\documentstyle[12pt]{article}
 \def\g{\mbox{${\bf g}$}}
 \def\F{\mbox{${\cal F}$}}
\begin{document}
\baselineskip16pt
\title{ $\kappa$-deformations of $D=4$ Weyl and conformal 
symmetries\thanks{Supported by KBN grant No 5PO3B05620 (JL 
and MM) and by the
 Russian
 Foundation for Basic Research under grant No 00-01-00500 (VL).}}

\author{Jerzy Lukierski$^{1)}$, Vladymir Lyakhovsky$^{2)}$
 and Marek Mozrzymas$^{1)}$
\\
$^{1)}$Institute for Theoretical Physics, \\
 University of Wroc{\l}aw,                  \\
pl. Maxa Borna 9,                             \\
 50--205 Wroc{\l}aw, Poland
\\
$^{2)}$Departament of Theoretical Physics, Sankt-Petersburg 
University, \\
         Ulianovskaya 1, Petrodvoretz, \\
198904 Sankt Petersburg, Russia}

\begin{titlepage}
\def\thepage{}
\maketitle

\begin{abstract}
We provide first explicite examples of quantum deformations of
$D=4$ conformal algebra with mass-like deformation parameters, 
in 
applications to quantum gravity effects related with Planck mass. It
is shown that one of the classical $r$-matrices defined on the
Borel subalgebra of $sl(4)$ with $o(4,2)$ reality conditions
describes the light-cone $\kappa$-deformation of $D=4$
Poincar\'{e} algebra. We embed this deformation into the
three-parameter family of generalized $\kappa$-deformations, with
$r$-matrices depending additionally on the dilatation generator.
Using the extended Jordanian twists framework we describe these
deformations in the form of noncocommutative Hopf algebra. We
describe also another four-parameter class of generalized
$\kappa$-deformations, which is  obtained by continuous
deformation of distinguished $\kappa$-deformation of $D=4$ Weyl
algebra, called here the standard $\kappa$-deformation of Weyl
algebra.
\end{abstract}

\end{titlepage}
\def\thepage{\arabic{page}}
{
 \noindent {\sl {\bf I. Introduction. \ }}

There are two basic classes of deformations of $D=4$ relativistic
symmetries:

i) $q$-deformations (see e.g.[1-3]), with Drinfeld-Jimbo
deformations [4,5] of the Lorentz subalgebra. The basic relations
describing noncommutative space-time coordinates are quadratic
[6,7] and the deformation parameter is dimensionless. It can be
shown [8,9] that in order to obtain the  $q$-deformation of $D=4$
Poincar\'{e} algebra with Hopf algebra structure one should
introduce additional generators, i.e. consider  $11$-dimensional
quantum Weyl algebra $U_{q}(W)$ ($W\equiv$Poincar\'{e} algebra
$P^{3;1}\oplus$ dilatation generator $D$, where
$P^{3;1}=(P_{\mu}, M_{\mu \nu})$). There is a natural Hopf
algebra  embedding of the $q$-deformed Weyl algebra $U_{q}(W)$
into the Drinfeld-Jimbo deformation $U_{q}(o(4,2))$ of $D=4$
conformal algebra
\begin{equation}
U_{q}(o(4,2))\supset U_{q}(P^{3;1}\oplus D)
\end{equation}
The $q$-deformation of $D=4$ conformal algebra was studied by
several authors (se e.g. [10-13]).

ii) $\kappa$-deformations [14-21] with deformation parameter
$\kappa^{-1}$ introducing fundamental dimensionfull length
parameter \footnote{If we put $h=c=1$, the parameter
$\Lambda=\kappa^{-1}$ is the fundamental length parameter, and
$\kappa$ describes the fundamental mass. For undeformed case
$\kappa\rightarrow \infty$.}. The basic relations describing
$\kappa$-deformed space-time are Lie-algebraic, and in the general
case take the form
\begin{equation}
[\hat{x}_{\mu}, \hat{x}_{\nu}]=\frac{i}{\kappa}
(a_{\mu}\hat{x}_{\nu}-\hat{x}_{\mu}a_{\nu})
\end{equation}
where $a_{\mu}$ is a constant four-vector in Minkowski space,
determining the quantum direction $\hat{y}=a^{\mu}\hat{x}_{\mu}$
\footnote{The $\kappa$-deformation for arbitrary choice of the
four vector $a_{\mu}$ is equivalent to the description of
standard $\kappa$-deformation in space-time with arbitrary
symmetric metric tensor, firstly given in [20,21]}. In fact due to
the choice of the four-vector $a_{\mu}$ one obtains the following
three particular cases of $\kappa$-deformation:

ii1)$a_{\mu}a^{\mu}=1$ defining time-like $\kappa$-deformations.
The standard $\kappa$-defomation [17-19] obtained for
$a_{\mu}=(1,0,0,0), (i,j=1,2,3)$ leads to relations\\
\begin{equation}
[\hat{x}_{0},\hat{x}_{i}]=\frac{i}{\kappa}\hat{x}_{i}\qquad
[\hat{x}_{i},\hat{x}_{j}]=0
\end{equation}

ii2) $a_{\mu}a^{\mu}=-1$ defining tachyonic $\kappa$-deformations.
For $a_{\mu}=(0,1,0,0), (r,s=1,2,)$ one gets\\
\begin{eqnarray}
[\hat{x}_{3},\hat{x}_{0}]&=&\frac{i}{\kappa}\hat{x}_{0}\qquad
[\hat{x}_{3},\hat{x}_{r}]=\frac{i}{\kappa}\hat{x}_{r}\cr
[\hat{x}_{3},\hat{x}_{r}]&=&\frac{i}{\kappa}\hat{x}_{r}\qquad
[\hat{x}_{r},\hat{x}_{s}]=0
\end{eqnarray}
Such a deformation leads to the deformation of nonrelativistic
symmetries, with classical time parameter.

ii3)$a_{\mu}a^{\mu}=0$ providing light-cone $\kappa$-deformations.
For $a_{\mu}=\frac{1}{\sqrt{2}}(1,1,0,0)$ one obtains [20-22]
\begin{eqnarray}
[\hat{x}_{-},\hat{x}_{+}]&=&\frac{i}{\kappa}\hat{x}_{+}\qquad
[\hat{x}_{-},\hat{x}_{r}]=\frac{i}{\kappa}\hat{x}_{r}\cr
[\hat{x}_{-},\hat{x}_{r}]&=&0\qquad [\hat{x}_{r},\hat{x}_{s}]=0
\end{eqnarray}
where $\hat{x}_{\pm}=\frac{1}{\sqrt{2}}(\hat{x}_{0}\pm
\hat{x}_{3})$ ($x_{+}=\tilde{a}^{\mu}\hat{x}_{\mu}$, where
$\tilde{a}^{\mu}\tilde{a}_{\mu}=0$ and
$a_{\mu}\tilde{a}^{\mu}=1$). The first quantum deformation with
quantum-deformed direction on the light cone was proposed in [23]
under the name of the null-plane quantum Poincar\'{e} algebra. It
was shown [22] that after suitable choice of basis the quantum 
algebra 
presented in [23] can be
identified with choice ii3) of $a_{\mu}$-dependent
$\kappa$-deformation, given in [20].

It appears that one can introduce the quantum deformation of
Poincar\'{e} group $P_{\kappa}^{3;1}(a_{\mu})$ with the
translation sector described by the relations (2) and the
corresponding dual quantum Poincar\'{e} algebra
$U_{\kappa}(P^{3;1};a_{\mu})$ [20,21]. One can show\footnote{For
an arbitrary $a_{\mu}$ the classical $\hat{r}$-matrix is given in
[20], where $g_{\mu\nu}=\left(
\begin{array}{cccc}0&1&0&0\\
1&0&0&0\\
0&0&1&0\\
0&0&0&1\\
\end{array} \right)$ corresponds to the light-like choice of
$a_{\mu}$.} that only in the case ii3), for the light-cone
$\kappa$-deformation, the corresponding classical $r$-matrix
$r\in P^{3;1}\bigotimes P^{3;1}$ satisfies the classical Y-B
equation\footnote{This result shown in [20] is also contained
implicitely in [23].}. For the choice of the four-vector
$a_{\mu}=\frac{1}{\sqrt{2}}(1,1,0,0)$ it takes the form
\begin{equation}
r=\frac{1}{2}(L_{3}\wedge P_{+}-(M_{1}-L_{2})\wedge
P_{2}+(M_{2}+L_{1})\wedge P_{1})
\end{equation}
where $M_{i}$ describe three space rotations, $L_{i}$ represent
three Lorentz boosts and $P_{\pm}=P_{0}\pm P_{3}$. The
Poincar\`{e} algebra $P^{3;1}$ is the subalgebra of $D=4$ Weyl
algebra $W$ and  $D=4$ conformal algebra $C^{3;1}\simeq o(4,2)$,
therefore the classical $r$-matrix (6) is an example of the
solution of the classical Y-B equation for $D=4$ Weyl and
conformal algebras. It follows therefore that the $r$-matrix (6)
 describes
infinitesimally the quantum deformation $U_{\kappa}(W)$ of $D=4$
Weyl algebra (see [20]) as well as the  quantum deformation
$U_{\kappa}(o(4,2))$ of $D=4$ conformal algebra.

The aim of this paper is to describe the large family of
generalized $\kappa$-deformations of $D=4$ Weyl and conformal
algebras in the form of Hopf algebras. One can show that the
classical $r$-matrix (6) is defined on the Borel subalgebra
$B_{+}(o(4,2))$ of the Lie algebra $o(4,2)$, described as the
real form of the complex Lie algebra $sl(4)$. In section 2 we
consider the Cartan-Weyl basis for $o(4,2)$ algebra and a general
family of classical $\hat{r}$-matrices satisfying CYBE and
belonging to the tensor square of $B_{+}(o(4,2))$. We obtain two
classes of $r$-matrices, with three and four arbitrary
parameters, all introducing deformation parameters with the
dimension of inverse mass \footnote{We shall call these
deformations the generalized $\kappa$-deformations, in accordance
with our dychotomic split of quantum deformations of space-time
symmetries into $q$-deformations (dimensionless) and
$\kappa$-deformations (dimesionfull).}. We shall quantize all
these solutions of CYBE by generalized extended Jordanian twists,
which define classical universal enveloping $D=4$ Weyl and
conformal algebras with twisted coproducts. The physical $D=4$
conformal basis and the expressions in terms of conformal basis
of extended Jordanian twists [25-27] are given in section 3. A
new basic $\kappa$-deformation of $D=4$ Weyl algebra will be
considered in some detail. In section 4 we shall present comments
on possible physical applications.

We would like to mention that our paper extends to the $D=4$ case
the results known already for $D=1, D=2$ and $D=3$. For $D=1$ it
was shown in [28] that the so-called Jordanian deformation of
$sl(2;C)$ [29,30] by imposing the real form selecting
$sl(2,R)\simeq o(2,1)$ generators describes the
$\kappa$-deformation of $D=1$ conformal algebra. Because for
$D=2$ conformal algebra $O(2,2)\simeq O_{+}(2,1)\bigoplus
O_{-}(2,1)$, the results in [28] do extend in straight-forward
way, with  a pair of $D=1$ $\kappa$-deformations for two
$O_{+}(2,1)$ and $O_{-}(2,1)$ subalgebras described by two
fundamental mass parameters $\kappa_{\pm}$. An example of
$\kappa$-deformation of $D=3$ conformal algebra $o(3,2)$ has 
been
given by Herranz [31], which can be treated as
$\kappa$-deformation od $D=4$ anti-de-Sitter algebra, with two
dimensionfull parameters: the first, $\kappa$, possibly equal to
Planck mass, related with quantum gravity effects at ultra-short
distances and the second, represented by anti-de-Sitter radius $R$,
characterizing very large cosmological distances \footnote{ One
can show that the  $\kappa$-deformation of $D=4$ anti-de-Sitter
algebra considered in [31] provides in the contraction limit
$R\rightarrow\infty$  the light-cone $\kappa$-deformation of
Poincar$\acute{e}$ algebra, with classical $r$-matrix (6).}. All
these deformations of $D<4$ conformal algebras are generated by
classical $r$-matrices having support in corresponding Weyl
algebras. In this paper for the case $D=4$ we describe also the
multiparameter $\kappa$-deformations of conformal algebras as
induced by the
$\kappa$-deformations of Weyl algebra .\\

%

\noindent {\sl {\bf II. Classical $D=4$ conformal $r$-matrices and
$\kappa$--deformations. \ }}\\
 a) \textsl{$sl(4;c)\cong so(6;c)$ Lie
algebra in the Cartan-Weyl basis.}

The Cartan-Weyl basis for $sl(4)$ is defined by the simple root
generators $e_{\pm i}$, ($i=1,2,3$) and the composite generators:
\begin{eqnarray}
e_{4}= [e_{1},e_{2}],\qquad e_{-4}= [e_{-2},e_{-1}],\\
e_{5}= [e_{2},e_{3}],\qquad e_{-5}= [e_{-3},e_{-2}],\\
e_{6}= [e_{1},e_{5}],\qquad e_{-6}= [e_{-5},e_{-1}].
\end{eqnarray}
Let $h_{i}$ ($i=1,2,3$) form the linear basis of the Cartan
subalgebra and introduce the elements
\begin{equation}
h_{4}  =  h_{1}+h_{2},\quad h_{5}=h_{2}+h_{3},\quad h_{6}
=h_{1}+h_{2}+h_{3}.
\end{equation}
The Lie algebra $sl(4)$  can be completely described as follows:
\begin{equation}
[h_{A},e_{\pm B}]  = \pm \alpha_{AB}e_{\pm B},\qquad
[e_{A},e_{-B}]  = \delta_{AB}h_{B},
\end{equation}
($A=1,2,...,6$; no summation over repeated indices!). Here
$\alpha_{AB}$ is the extended symmetric Cartan matrix  given by
the formula (see e.g.[13])
\begin{equation}
\alpha_{AB}=\left(
\begin{array}{cccccc}2&-1&0&1&-1&1\\
-1&2&-1&1&1&0\\
0&-1&2&-1&1&1\\
1&1&-1&2&0&1\\
-1&1&1&0&2&1\\
1&0&1&1&1&2\end{array} \right)
\end{equation}
The remaining part of $sl(4)$ Lie algebra is obtained from the
Serre relations and for the positive root generators it takes the
form:
\begin{eqnarray}
[e_{1},e_{3}]&=&[e_{1},e_{4}]=[e_{1},e_{6}]=0,\cr
[e_{2},e_{4}]&=&[e_{2},e_{5}]=[e_{2},e_{6}]=0,\cr
[e_{4},e_{5}]&=&[e_{4},e_{6}]=[e_{5},e_{6}]=0,\cr
[e_{3},e_{5}]&=& [e_{3},e_{6}]=0,\qquad [e_{3},e_{4}]=e_{6}.
\end{eqnarray}

The real forms of $sl(4;c)$ are well-known (see e.g.[13,33]). We
shall consider the one which selects the $D=4$ real conformal
algebra ($sl(4)\rightarrow su(2,2)\simeq o(4,2)$) and map the
Borel subalgebras $B_{\pm}(o(4,2))\longrightarrow
B_{\pm}(o(4,2))$. We obtain
\begin{eqnarray}
h_{1}^{+}&=&-h_{3}, \qquad\qquad h_{2}^{+}=-h_{2},\cr
e_{1}^{+}&=&\epsilon e_{3}, \quad\qquad\qquad e_{2}^{+}=-e_{2},\cr
e_{4}^{+}&=&-\epsilon e_{5}, \qquad\qquad e_{6}^{+}=-e_{6},
\end{eqnarray}
where $\epsilon=\pm1$.\\

\noindent b)\textsl{Classical $r$-matrices for $o(4,2)$}.

We shall consider the following class of  classical
eight-dimensional $r$-matrices\footnote{Following [33,34] one can
introduce [32] the classical $r$-matrices for $sl(4)$ satisfying
CYBE with the support in $10$-dimensional and $12$-dimensional
parabolic subalgebras, but these $r$-matrices contain the
generators from both Borel algebras and are not compatible with
$o(4,2)$ reality conditions (14).}for $sl(4,c)$ having their
support in the Borel subalgebra $B_{+}$:
\begin{equation}
r_{+}=H_{2}\wedge e_{2}+H_{6}\wedge e_{6}+ce_{2}\wedge e_{6}+
+e_{1}\wedge e_{5}-e_{3}\wedge e_{4}
\end{equation}
where $H_{2}$ and $H_{6}$ are the  linear combinations of Cartan
generators with complex coefficients
\begin{equation}
H_{2}=b_{1}h_{1}+b_{2}h_{2}+b_{3}h_{3},\qquad
H_{6}=a_{1}h_{1}+a_{2}h_{2}+a_{3}h_{3}.
\end{equation}
We found the following two families of solutions for CYBE:

i) Let $H_{2} \neq 0$. In this case one gets the four-parameter
family of classical $r$-matrices, obtained by imposing three
conditions on seven parameters $a_{i}, b_{i}, c$ (i=1,2,3):
\begin{equation}
a_{1}+a_{3}=1,\qquad b_{1}+b_{3}=0,\qquad a_{2}=\frac{1}{2}.
\end{equation}

ii) If $H_{2}=0$ one obtains only one condition
\begin{equation}
a_{1}+a_{3}=1,
\end{equation}
i.e. the three-parameter set of $r$-matrices.

In order to obtain the classical $r$-matrices for $o(4,2)$ one
should impose the reality conditions $r_{+}=(r_{+})^{+} $ (notice
that  ($A\otimes B)^{+}=A^{+}\otimes B^{+}$) by the relations
(14). One obtains the following conditions on seven complex
parameters $a_{i}, b_{i}, (i=1,2,3), c$
\begin{equation}
a_{1}=a_{3}^{\ast},\quad a_{2}=a_{2}^{\ast},\qquad
b_{1}=b_{3}^{\ast},\quad b_{2}=b_{2}^{\ast},\qquad c=c^{\ast}.
\end{equation}

Imposing the conditions (17) one obtains the first four-parameter
family of classical $o(4,2)$ $r$-matrices ($\alpha, \beta,
\gamma, \delta$ real)
\begin{equation}
r_{+}^{(1)}(\alpha, \beta, \gamma, \delta)=i(h_{1}-h_{3})\wedge
(\alpha e_{2}+\beta e_{6})+ \delta h_{2}\wedge e_{2}+\gamma
e_{2}\wedge e_{6}+\hat{r}
\end{equation}
where $\hat{r}\equiv r_{+}^{(1)}(0, 0, 0,0)$ has the form
\begin{equation}
\hat{r}=r+\frac{1}{2}h_{2}\wedge e_{6}
=\frac{1}{2}(h_{1}+h_{2}+h_{3})\wedge e_{6}+e_{1}\wedge
e_{5}-e_{3}\wedge e_{4}
\end{equation}
We shall show that the generators occuring in (20) belong to
$D=4$ Weyl algebra and will be called the standard $\kappa$-Weyl
$r$-matrix. It should be pointed out that the $D=1$
$\kappa$-deformation considered in [28] and $D=3$
$\kappa$-deformation from [31] are of such a type.

The second class of $o(4,2)$ $r$-matrices is obtained by imposing
the conditions (18,19). One gets ($\beta, \gamma, \rho$ real)
\begin{equation}
r_{+}^{(2)}(\beta, \gamma, \rho)=i\beta(h_{1}-h_{3})\wedge e_{6}+
\rho h_{2}\wedge e_{6}+\gamma e_{2}\wedge e_{6}+r
\end{equation}
where $r\equiv r_{+}^{(2)}(0, 0, 0)$ is  the light-cone
$\kappa$-Poincar\`{e} $r$-matrix (6) in the Cartan-Weyl basis
\footnote{This statement follows from formulae (37) in section 3.}
\begin{equation}
r=\frac{1}{2}(h_{1}+h_{3})\wedge e_{6}+e_{1}\wedge
e_{5}-e_{3}\wedge e_{4}
\end{equation}
The set $r_{+}^{(2)}(\beta, \gamma, \rho)$ contains the
$r$-matrix (21) as a limit: $r_{+}^{(2)}(0, 0,
\frac{1}{2})=\hat{r}$. To avoid this reduction we shall assume
that $\rho\neq\frac{1}{2}$.

To see what parameters in classical $r$-matrices are essential
one should consider the group of outer and inner automorphisms of
the Borel subalgebra $B_{+}$, belonging to $O(4,2)$. Applying the
automorphism generated by $e_{2}$ to the classical $r$-matrices
$r_{+}^{(2)}$ (see (22)) one gets
\begin{equation}
exp[ad^{\otimes}(\alpha_{2}e_{2})]r_{+}^{(2)}=
r_{+}^{(2)}+\alpha_{2}(1-2\rho)e_{2}\wedge e_{6}
\end{equation}
i.e. the parameter $\gamma$ is shifted ($\gamma \rightarrow
\gamma'=\gamma + \alpha_{2}(1-2\rho)$) and while
$\rho\neq\frac{1}{2}$ it can be removed. One can check that do
not exist the  $O(4,2)$ automorphisms which  can be used to
eliminate or  scale independently the parameters $\beta$ and
$\rho$. Similarly it can be shown that $O(4,2)$ automorphisms can
not eliminate  any of four parameters $\alpha, \beta, \gamma,
\delta$ in (20). We remind that for matrices $r_{+}^{(1)}$ we
have $\rho =\frac{1}{2}$ and parameter $\gamma$  can not be
eliminated through shift (24), however it can be  scaled.\\
\noindent c)\textsl{Quantum deformations by twisting classical Lie
algebras.}

Consider the universal enveloping algebra $U(\g)$ of a Lie algebra
$\g$ as a Hopf algebra with the comultiplication $\Delta^{(0)}$
generated by the primitive coproducts of generators in $\g$. The
parametric solution $\F(\xi) = \sum f^{(1)}_{i} \otimes
f^{(2)}_{i} \in U(\g)\otimes U(\g)$ of the twist equations [35]
\begin{eqnarray}
\F_{12}(\Delta^{(0)}\otimes 1)\F
=\F_{23}(1\otimes\Delta^{(0)})\F, \\
(\epsilon \otimes {\rm id}) \F = ({\rm id} \otimes \epsilon)\F = 1
\otimes 1.
\end{eqnarray}
defines the deformed (twisted) Hopf algebra  $U_{\F}(\g)$ with the
unchanged multiplication, unit and counit (as in  $U(\g)$), the
twisted comultiplication and antipode defined by the relations
\begin{eqnarray}
\Delta_{\F}(u)&=&\F \Delta^{(0)}(u)\F^{-1}, \quad u \in U(\g), \cr
 S_{\F}(u)&=&-v u v^{-1}, \quad
v=\sum f^{(1)}_{i}(S^{(0)}f^{(2)}_{i}).
\end{eqnarray}
The twisted algebra  $U_{\F}(\g)$ is triangular, with the
universal ${\cal R}$-matrix
\begin{equation}
{\cal R}_{\F}=\F_{21}\F^{-1}.
\end{equation}
When $\F$ is a smooth function of $\xi$ and $\lim_{\xi
\rightarrow 0} \F= 1 \otimes 1$ then in the neighborhood of the
origin the ${\cal R}$-matrix can be presented as
\begin{equation}
{\cal R}_{\F}= 1 \otimes 1 + \xi r_{\F} + o(\xi),
\end{equation}
where $ r_{\F}$ is the skewsymmetric classical $r$-matrix
corresponding to the twist $\F$.
 
By a nonlinear change of basis in $\g$ one can modify the twisted
coproducts and locate  part of the deformation in the algebraic
sector.\\
\noindent d)\textsl{Twisting elements and real forms of Hopf
algebras}

The generators of physical symmetries are described by
self-adjoint (real) generators and it is desirable to have a real
form of quantized Lie algebras. Real forms of complex simple Lie
algebra $g$ are described by the involutive automorphism $\Phi:
U(\hat{g})\rightarrow U(\hat{g})$ with the following properties
\begin{eqnarray}
\Phi^{2}&=&{\rm id}, \qquad \Phi ({\bf 1}) = {\bf 1},\cr
\Phi(X\cdot Y)&=&\Phi(Y)\cdot\Phi(X),\cr \Phi(\mu X+\nu
Y)&=&\mu^{\ast}\Phi(X)+\nu^{\ast}\Phi(Y).
\end{eqnarray}
The Hopf algebra $(U(\hat{g}), \cdot,
\Delta,S,\eta,\varepsilon,\Phi )$ that in general has
nonprimitive costructure is a standard real Hopf algebra if $\Phi$
satisfies (30) and additionally
\begin{equation}
\Delta(\Phi(X))=\Phi(\Delta(X))
\end{equation}
where $\Phi(X\otimes Y)=\Phi(X)\otimes \Phi(Y)$. This implies
\begin{equation}
(\Phi \circ S)^{2}=1.
\end{equation}
In particular for $\Delta_{\F}$ given by (27) one gets
\begin{equation}
\Phi(\Delta_{\F}(X))=\Phi(\F^{-1}) \Delta^{(0)}(\Phi(X))\Phi(\F),
\end{equation}
where we used the relation (31) for $\Delta = \Delta^{0}$ (the
comultiplication induced by primitive coproducts of generators).
From (33) one gets (instead of (31))
\begin{equation}
\Delta_{\F}(\Phi(X))=\F \Delta^{(0)}(\Phi(X))\F^{-1}=J
\Phi(\Delta_{\F}(X))J^{-1},
\end{equation}
where
\begin{equation}
 \F\cdot\Phi(\F)=J.
\end{equation}
Therefore we see that the standard definition of real form for
quantum group is valid only if the twisting element $\F$ is
"$\Phi$-unitary", i.e. $[\Delta_{\F}(X),J]=0$ for any $X \in
U(\hat{g})$. In such a case one can say that the twist $\F$ is
compatible with the involution $\Phi$ (with the relation (30)
satisfied). Due to the formulae (35) the element $J$ is
"$\Phi$-unitary", i.e.
$[J\cdot\Phi(J^{-1}),\Delta_{\F}(X)]=0$.

It is known that for $sl(n;c)$ the extended Jordanian twists are
naturally compatible with the involution providing the real form
$sl(n;R)$ (see [25], sec. 5). In the general case the involutions
defining other real forms are not compatible with extended
Jordanian twists.
\\[24pt]
\noindent e)\textsl{Twists for $\kappa$-deformed conformal
algebra}

The general approach to the solution of twist equations for the
Lie algebra $sl(n)$  was presented in \cite{lun16a,lun16c}. Here
we shall apply these results for the case $n=4$.

The two-parameter family of classical $r$-matrices (22) that
contains the light-cone $\kappa$-Poincar\`{e} $r$-matrix (we
recall that in (22) one can put $\gamma=0$) leads to the
following three-parameter twisting element:
 \begin{equation}
\begin{array}{rcl}
\F_{\xi}(\beta,  \varrho)& =  & exp[\xi e_{1}\otimes
e_{5}(1+\xi e_{6})^{(2i\beta-\varrho)}]\\
& & exp[\xi
e_{4}\otimes e_{3}(1+\xi e_{6})^{(2i\beta+\varrho-1)}]\\
& & exp{[i\beta(h_{1}-h_{3})+\varrho
h_{2}+\frac{1}{2}(h_{1}+h_{3})]\otimes \sigma_{6}(\xi)]}
\end{array}
\end{equation}
where (we recall that $\rho\neq \frac{1}{2}$)
\begin{equation}
\sigma_{k}=\ln(1+\xi e_{k}).
\end{equation}
and the reality condition $\xi^{\ast}=\pm \xi$ copies the reality
condition for $e_{k}$ (see (14)) i.e. $(\xi e_{k})^{\ast}=\xi
e_{k}$ , therefore in (37) $\xi^{\ast}=- \xi$. In this expression
$\xi$ plays the role of an "over-all" deformation parameter and
provides the expansion
\begin{equation}
\F_{\xi}(\beta,  \varrho)= 1+\xi r^{ns}(\beta,
\varrho)+O(\xi^{2}).
\end{equation}
Where the  element $r^{ns} \in \g \otimes \g$ can be obtained from
(22) by substituting the wedge product $(A\wedge B = A\otimes B -
B\otimes A)$ by the tensor product: $(A\wedge B\rightarrow
A\otimes B)$ and setting $\gamma=0$.  If we assume that in (36)
$\beta=\varrho =0$ we obtain the formula for the twisting
element  $\F_{\xi}$ for the light-cone $\kappa$-Poincar\`{e}
deformation
\begin{equation}
\begin{array}{rcl}
\F_{\xi} & \equiv & \F_{\xi}(0,0))=\\
& & exp(\xi e_{1}\otimes e_{5})exp[\xi e_{4}\otimes e_{3}(1+\xi
e_{6})^{-1}]exp[\frac{1}{2}(h_{1}+h_{3})\otimes \sigma_{6}(\xi)].
\end{array}
\end{equation}

The four-parameter family of infinitesimal deformations described
by classical $r$-matrices (20) leads to the  the following
six-parameter family of twists
\begin{equation}
\begin{array}{rcl}
\hat{\F}_{\xi,\xi'}(\alpha, \beta, \gamma, \delta)& = &
exp[\frac{\gamma}{\xi'}\sigma_{2}(\xi')\otimes \sigma_{6}(\xi)
]exp[(\frac{i\alpha}{2\delta}(h_{1}-h_{2})+\frac{1}{2}h_{2})
\otimes\sigma_{2}(\xi')]\\
& & exp[\xi e_{1}\otimes e_{5}(1+\xi e_{6})^{(2i\beta-\frac{1}{2})}]\\
& &  exp[\xi e_{4}\otimes e_{3}(1+\xi e_{6})^{(2i\beta-\frac{1}{2})}]\\
& & exp{[i\beta(h_{1}-h_{3})+\frac{1}{2}(h_{1}+h_{2}+h_{3})]
\otimes \sigma_{6}(\xi)}.
\end{array}
\end{equation}
where $\xi^{\ast}=-\xi,\quad (\xi')^{\ast}=-\xi'$. The first
factor in (40) contributes to the classical $r$-matrix due to the
relation
\begin{equation}
exp[\frac{\gamma}{\xi'}\sigma_{6}(\xi)\otimes
\sigma_{2}(\xi')]=exp \gamma\xi (e_{6}-\xi
e_{6}^{2}/2+O(\xi^{2}))\otimes (e_{2}-\xi'
e_{2}^{2}/2+O(\xi'^{2})).
\end{equation}

Both parametric families of twists (36) and (40) start with the
similar extended Jordanian twists. The difference between the
families $\F_{\xi}$ and $\hat{\F}_{\xi,\xi'}$ is that the second
of them is the full chain of twists containing two links with the
additional Reshetikin factor while the first is the single
extended twist [25,26]. The intersection of these families is the
two-parametric set  $\F_{\xi}(\beta, 1/2)$.

Due to the "matreshka" effect \cite{lun16c} the deformation
parameters $\xi$ and $\xi'$ are independent. To pass to the
quasi-classical limit one must choose in the space of parameters
the one-dimensional smooth curve. We shall consider here the case
where $\xi'$ is proportional to $\xi$. In particular for
$\xi'=2\delta \xi$ and $\xi$ as the overall deformation parameter
we obtain the classical $r$-matrix (20) as the leading term in
the expansion of the universal $R$-matrix ${\cal
R}(\xi)=(\hat{\F}_{\xi,\xi})_{21}(\hat{\F}_{\xi,\xi})^{-1}$.\\
 The
twist $\hat{\F}_{\xi} \equiv \F_{\xi}(0,0,0,0)$ for $\kappa$-Weyl
deformation described by the classical $r$-matrix $\hat{r}$ (see
(20)) is obtained from (40) at $\beta =0$ at particular limits
$\gamma\rightarrow 0 ,\xi'\rightarrow 0$ as well as
$\alpha\rightarrow 0, \delta\rightarrow 0$ and is given by the
formula
 \begin{equation}
\hat{\F}_{\xi}=exp[\xi e_{1}\otimes e_{5}(1+\xi
e_{6})^{-\frac{1}{2}}]exp[\xi e_{4}\otimes e_{3}(1+\xi
e_{6})^{-\frac{1}{2}}]exp[\frac{1}{2}(h_{1}+h_{2}+h_{3})]\otimes
\sigma_{6}(\xi)]
\end{equation}

\noindent {\sl {\bf III. $D=4$ conformal $\kappa$-deformations
described by
  twists. \ }}

i) \textsl{Physical basis for $D=4$ conformal algebra  and
classical
  conformal $r$-matrices}.\\
The $D=4$ conformal Lie algebra is described by 15 anti-hermitean
generators $J_{AB}$ ($J_{AB}^{+}=-J_{AB}=J_{BA};
A,B=0,1,2,3,4,5$) satisfying the relation
\begin{equation}
[J_{AB},J_{CD}]=\eta_{AD}J_{BC}+\eta_{BC}J_{AD}-\eta_{BD}
J_{AC}-\eta_{AC}J_{BD}
\end{equation}
with $\eta_{AB}=diag(-1,1,1,1,-1)$. The physical basis is given
by the relations
\begin{eqnarray}
P_{\mu}&=&\frac{1}{\sqrt{2}}(M_{5\mu}+M_{4\mu}), \qquad K_{\mu} 
=
\frac{1}{\sqrt{2}}(M_{5\mu}-M_{4\mu}),\cr
M_{i}&=&\frac{1}{2}\epsilon_{ijk}J_{jk},\qquad
L_{i}=M_{0i},\qquad D=M_{45}
\end{eqnarray}
where $\mu=0,1,2,3$ and $i,j=1,2,3$. All the generators (44) are
also anti-hermitean. For convenience we introduce also the the
generators $P_{\pm}= P_{0}\pm P_{3}$ as well as 
$M_{\pm}=M_{1}\pm
iM_{2}\quad ((M_{\pm})^{+}=-M_{\mp})$ and $L_{\pm}=L_{1}\pm 
iL_{2}
\quad((L_{\pm})^{+}=-L_{\mp})$.
 The Cartan-Weyl
basis is described in terms of $D=4$ conformal generators ($M_{i},
N_{i}, D, P_{i}, P_{0}, K_{i}, K_{0}$) as follows
\begin{eqnarray}
e_{1}&=&\frac{1}{2}(M_{+}+iL_{+}),\qquad
e_{-1}=\frac{1}{2}(-M_{-}-iL_{-}),\cr
e_{2}&=&\frac{1}{2}P_{-},\qquad e_{-2}=-\frac{1}{2}K_{+},\cr
e_{3}&=&-\frac{1}{2}(M_{-}-iL_{-}),\qquad
e_{-3}=\frac{1}{2}(M_{+}-iL_{+}),\cr
e_{4}&=&\frac{i}{2}(P_{1}+iP_{2}),\qquad
e_{-4}=-\frac{i}{2}(K_{1}-iK_{2}),\cr
e_{5}&=&-\frac{i}{2}(P_{1}-iP_{2}),\qquad
e_{-5}=\frac{i}{2}(K_{1}+iK_{2}),\cr
e_{6}&=&\frac{1}{2}P_{+}\qquad e_{-6}=-\frac{1}{2}K_{-},\cr
h_{1}&=& L_{3}-iM_{3},\qquad h_{2}=- L_{3}-D, \qquad h_{3}=
L_{3}+iM_{3}.
\end{eqnarray}

The classical $r$-matrix  (23) describing light-like
$\kappa$-deformation is given respectively by the formula (6),
with the "overall" deformation parameter $\xi=\frac{2i}{\kappa}$
carrying the inverse mass dimension, i.e. we obtain the following
expansion for universal $R$-matrices
\begin{equation}
R_{\kappa}=1+\frac{1}{\kappa}(L_{3}\wedge
P_{+}-(M_{1}-L_{2})\wedge P_{2}+(M_{2}+L_{1})\wedge
P_{1})+O(\frac{1}{\kappa^{2}})
\end{equation}
Similarly, the $\kappa$-Weyl $r$-matrix (21) provides the
expansion
\begin{equation}
\hat{R}_{\kappa}=1+\frac{1}{\kappa}[\frac{1}{2}(L_{3}-D)\wedge
P_{+}-(M_{1}-L_{2})\wedge P_{2}+(M_{2}+L_{1})\wedge
P_{1}]+O(\frac{1}{\kappa^{2}})
\end{equation}
where the term linear in $\frac{1}{\kappa}$ describes the
$\kappa$-Weyl $r$-matrix in physical basis (45).\\
Denoting
\begin{equation}
\tilde{E}_{1}=L_{1}+M_{2},\qquad
\tilde{E}_{2}=L_{2}-M_{1},\qquad\tilde{E}_{3}=L_{3}
\end{equation}
one gets for the light-cone $\kappa$-Poincar\`{e} $r$-matrix the
following six-dimensional carrier algebra
\begin{equation}
[\tilde{E}_{1},\tilde{E_{2}}]=0,\qquad
[\tilde{E}_{3},\tilde{E_{r}}]=\tilde{E}_{r},
\end{equation}
\begin{equation}
[\tilde{E}_{r},P_{+}]=0,\quad
[\tilde{E}_{r},P_{s}]=\delta_{rs}P_{+},\quad
[\tilde{E}_{3},P_{+}]=P_{+},\quad [\tilde{E}_{3},P_{s}]=0
\end{equation}
where $r,s=1,2$. If we introduce the notation
\begin{equation}
\hat{E}_{1}=L_{1}+M_{2},\qquad
\hat{E}_{2}=L_{2}-M_{1},\qquad\hat{E}_{3}=\frac{1}{2}(L_{3}-D)
\end{equation}
the carrier algebra for the $\kappa$-Weyl $r$-matrix  is the
following \\
\begin{equation}
[\hat{E}_{1},\hat{E_{2}}]=0,\qquad
[\hat{E}_{3},\hat{E_{r}}]=\hat{E}_{r},
\end{equation}
\begin{equation}
[\hat{E}_{r},P_{+}]=0,\quad
[\hat{E}_{r},P_{s}]=\delta_{rs}P_{+},\quad
[\hat{E}_{3},P_{+}]=P_{+},\quad
[\hat{E}_{3},P_{s}]=\frac{1}{2}P_{s}
\end{equation}
where $r,s=1,2$.

It follows from relations (45) that in the formulae (21) and (23)
the parameters $\alpha,\beta, \delta, \varrho$ have the dimension
of inverse mass and the parameter $\gamma$, occuring only in
(20), describes the "soft deformation" with the dimension of the
inverse mass square. We see therefore that all deformations of
conformal algebra considered in this paper are the ones with
dimensionfull deformation parameters i.e. describe  generalized
$\kappa$-deformations (see footnote 5).

In order to characterize physically the deformations provided by
the twists (39) and (42) we shall consider the twisted coproducts
for the four-momentum generators. Using the twist (36) one
obtains the following formulae:
\begin{eqnarray}
\Delta_{\F_{\xi}}(e_{6})&=&e_{6}\otimes
e^{\sigma_{6}(\xi)}+1\otimes e_{6},\cr
\Delta_{\F_{\xi}}(e_{2})&=&e_{2}\otimes
e^{(2\varrho-1)\sigma_{6}(\xi)}+1\otimes e_{2},\cr
\Delta_{\F_{\xi}}(e_{4})&=&e_{4}\otimes
e^{(2i\beta+\varrho-1)\sigma_{6}(\xi)}+1\otimes e_{4},\cr
\Delta_{\F_{\xi}}(e_{5})&=&e_{5}\otimes
e^{(-2i\beta+\varrho)\sigma_{6}(\xi)}+e^{\sigma_{6}(\xi)}\otimes
e_{5},
\end{eqnarray}
If we consider the light-cone $\kappa$-Poincar\`{e} deformation
(see (46))  we observe from (54) that after the substitution (45)
all the four-momentum components in Poincar\`{e} algebra basis
are endoved with nonprimitive coproducts. In order to obtain
primitive coproduct for the light-cone energy $P_{+}'$ we should
introduce the following nonlinear transformation of
$P_{+}$-variable\footnote{It is interesting to observe that the
relation (55) coincides with the deformation map describing
$\kappa$-deformed Weyl algebra in [36] (see formula (14)). We
thank Piotr Kosi$\acute{n}$ski for providing this observation.}:
\begin{equation}
\frac{1}{2}\xi P_{+}'=\sigma_{6}(\xi)=\ln(1+\frac{1}{2}\xi P_{+})
\end{equation}
Substituting $\xi=\frac{2i}{\kappa}$ one gets for the light-cone
$\kappa$-deformation  the following four-momentum coproducts:
\begin{eqnarray}
\Delta_{\F_{\xi}}(P_{+}')&=&P_{+}'\otimes 1+1\otimes P_{+}',\cr
\Delta_{\F_{\xi}}(P_{-})&=&P_{-}\otimes
e^{-\frac{iP_{+}}{\kappa}}+1\otimes P_{-},\cr
\Delta_{\F_{\xi}}(P_{1}+iP_{2})&=&(P_{1}+iP_{2})\otimes
e^{-\frac{iP_{+}}{\kappa}}+1\otimes (P_{1}+iP_{2}),\cr
\Delta_{\F_{\xi}}(P_{1}-iP_{2})&=&(P_{1}-iP_{2})\otimes
1+e^{\frac{iP_{+}}{\kappa}}\otimes (P_{1}-iP_{2}),
\end{eqnarray}
The new basis with light-cone energy generator given by (55) leads
to the deformation of the commutators between Abelian
four-momentum generators ($P_{1}, P_{2}, P_{-}, P_{+}'$) and
classical Lorentz generators $M_{\mu\nu}=(M_{i}, N_{i})$:
\begin{eqnarray}
[M_{1},P_{+}']&=& P_{2}e^{-\frac{iP_{+}'}{\kappa}},\qquad
[M_{2},P_{+}']=- P_{1}e^{-\frac{iP_{+}'}{\kappa}},\cr
[M_{3},P_{+}']&=&0,\cr [L_{1},P_{+}']&=&
P_{1}e^{-\frac{iP_{+}'}{\kappa}},\qquad [L_{2},P_{+}']=
P_{2}e^{-\frac{iP_{+}'}{\kappa}},\cr
[L_{3},P_{+}']&=&-i\kappa(1-e^{-\frac{P_{+}'}{\kappa}}),
\end{eqnarray}
It should be noticed that the coproduct (56) for $P_{+}'$  remains
primitive for any choice of the parameters $ \beta$ and $\varrho$
 in the twist formulae (35).\\
It is easy to see that in terms of the  generators $P_{1}, P_{2},
P_{-}, P'_{+}$ the mass Casimir  takes the form
\begin{equation}
C_{2}=P_{1}^{2}+P_{2}^{2}-\frac{2}{\xi}(1-e^{\frac{1}{2}\xi
P'_{+}}) P_{-}
\end{equation}
 
The
coproduct formulae for the twist (40) which are analogous to the
ones given by the relations (54) look as follows:
\begin{eqnarray}
\Delta_{\F_{\xi,\xi'}^{w}}(e_{6})&=&e_{6}\otimes
e^{\sigma_{6}(\xi)}+1\otimes e_{6},\cr
\Delta_{\F_{\xi,\xi'}^{w}}(e_{2})&=&e_{2}\otimes
e^{\sigma_{6}(\xi')}+1\otimes e_{2},\cr
\Delta_{\F_{\xi,\xi'}^{w}}(e_{4})&=&e_{4}\otimes
e^{(2i\beta-
\frac{1}{2})\sigma_{6}(\xi)+(\frac{i\alpha}{\delta}+\frac{1}{2})\sigma_{2
}(\xi')}
+1\otimes e_{4},\cr
\Delta_{\F_{\xi,\xi'}^{w}}(e_{5})&=&e_{5}\otimes
e^{(\frac{1}{2}-2i\beta)\sigma_{6}(\xi)+(-
\frac{i\alpha}{\delta}+\frac{1}{2})\sigma_{2}(\xi')}+
e^{\sigma_{6}(\xi)}\otimes e_{5},
\end{eqnarray}
If we introduce  the expression (55) and redefine the energy
$P_{-}$
\begin{equation}
\frac{1}{2}\xi P_{-}'=\sigma_{2}(\xi)=\ln(1+\frac{1}{2}\xi P_{-})
\end{equation}
the  two coproducts, for the generators $P_{0}'$ and $P_{3}'$,
remain primitive. If $\alpha = \beta =0$ the formulae (59)
describe the coproducts
of four-momenta generated by the $\kappa$-Weyl twist (42). 

In order to obtain the quantum $D=4$ conformal algebra with more
suitable deformation of the algebra and coalgebra sectors  one
should at least supplement (55) and (60) with nonlinear
transformation of three-momenta (see also [36]). The choice of
"optimal" nonlinear basis in $U(o(4,2))$ describing the
$\kappa$-deformed $D=4$ conformal algebras $U_{\kappa}(o(4,2))$
is under consideration.\\

\noindent {\sl {\bf IV. Conclusions. \ }}

It is well-known that in conformal-invariant theories all
mass-like parameters should be put equal to zero; on the other
hand the Einstein gravity action is proportional to the
dimesionfull Newton coupling constant defining the Planck mass
$M_{PL}$ (or Planck lenght $\lambda_{PL}$). It is therefore
natural to assume that due to the gravitational effects in $D=4$
conformal-invariant field theory (e.g. non-Abelian gauge theory)
 should appear conformal-breaking terms, proportional to inverse
powers of the Planck mass. The $\kappa$-deformations of the
conformal symmetry algebra leads to a corresponding  algebraic
scheme of broken  conformal symmetries with the appearance of 
the
fundamental mass parameters, which physically should be related
with the Planck mass .

The advantage of our scheme is the description of broken Weyl and
conformal symmetries by exact quantum symmetries that permits 
to
use the algebraic techniques of quantum groups and quantum Lie
algebras. In particular a genuine (Hopf-algebraic) quantum
deformation of symmetry algebra permits to construct the
consistent representation theory for irreducible as well as for
tensor product representations, what should be useful in passing
from quantum-mechanical to quantum field-theoretic description of
models with broken
conformal symmetry.\\

\noindent {\sl Acknowledgments. \ } The authors would like to
thank Piotr Kosi$\acute{n}$ski for valuable comments.


\end{document}